\documentclass{nature}

\usepackage[utf8]{inputenc}

\usepackage{eso-pic}
\usepackage{graphicx}
\usepackage{color}
\usepackage{amsmath}
\usepackage{caption}
\usepackage{setspace}
\usepackage{ulem}
\usepackage{placeins}
\title{A metric for wettability at the nanoscale}

\author{Ronaldo Giro$^1$, Peter W. Bryant$^1$, Michael Engel$^2$, Rodrigo F. Neumann$^1$ \& Mathias Steiner$^{1,2}$}

\begin{document}

\maketitle

\begin{affiliations}
 \item IBM Research, Av. Pasteur 138/146, CEP 22290-240, Rio de Janeiro, RJ, Brazil
 \item IBM Research, 1101 Kitchawan Rd, Yorktown Heights, NY 10598, United States of America.
\end{affiliations}

\begin{abstract}

Wettability is the affinity of a liquid for a solid surface. For energetic reasons, macroscopic drops of liquid are nearly spherical away from interfaces with solids, and any local deformations due to molecular-scale surface interactions are negligible. Studies of wetting phenomena, therefore, typically assume that a liquid on a surface adopts the shape of a spherical cap. The degree of wettability is then captured by the contact angle where the liquid-vapor interface meets the solid-liquid interface\cite{de1985wetting,bonn2009wetting,yuan2013contact}.  As droplet volumes shrink to the scale of attoliters, however, surface interactions become significant, and droplets gradually assume distorted shapes that no longer comply with our conventional, macroscopic conception of a drop.  In this regime, the contact angle becomes ambiguous, and it  is unclear how to parametrize a liquid's affinity for a surface. A scalable metric for quantifying wettability is needed, especially given the emergence of technologies exploiting liquid-solid interactions at the nanoscale\cite{adera2013non,liu+14science, yin2014generating,bocquet2014nanofluidics, lee2014nanofluidic,giacomello2016wetting}. Here we combine nanoscale experiments with molecular-level simulation to study the breakdown of spherical droplet shapes at small length scales. We demonstrate how measured droplet topographies increasingly reveal non-spherical features as volumes shrink, in agreement with theoretical predictions. Ultimately, the nanoscale liquid flattens out to form layer-like molecular assemblies, instead of droplets, at the solid surface. For the lack of a consistent contact angle at small scales, we introduce a droplet's adsorption energy density as a new metric for a liquid's affinity for a surface.  We discover that extrapolating the macroscopic idealization of a drop to the nanoscale, though it does not geometrically resemble a realistic droplet, can nonetheless recover its adsorption energy if line tension is properly included.

\end{abstract}

According to our daily visual experience, macroscopic amounts of a liquid on a solid surface, surrounded by air, may assume a variety of shapes, perhaps resembling a shallow puddle with a complicated perimeter\cite{AlizadehPahlavan2015}. As the volume of liquid deposits decrease and gravitational influence becomes negligible, a drop's shape will be largely determined by minimizing the energetic contribution of the liquid's surface tension at the liquid-air interface, and thus appear spherical away from local distortions that arise from interactions at the liquid-solid interface.  As droplet dimensions approach the nanometer scale, surface-to-volume ratios increase and molecular-level phenomena at the solid surface become energetically significant. In this regime droplets' shapes are strongly influenced by effects such as line tension\cite{Amirfazli2004}, Tolman lengths\cite{Tolman1949}, and precursor film formation\cite{AlizadehPahlavan2015}. Though nanoscale droplets have non-trivial shapes, earlier studies have approximated them by spherical caps\cite{Checco2003,Berg2010}, thus extending the contact angle framework from the macroscale to the nanoscale where a suitable metric for wettability is lacking. Creating, isolating, and measuring nanoscale droplets is challenging, however, and purely geometrical figures of merit may suffer from inherent uncertainties associated with nanoscale measurements, see e.g. reference~\cite{Berg2010}.

\begin{figure*}[!htbp]
\centerline{\includegraphics[width=17.0 cm]{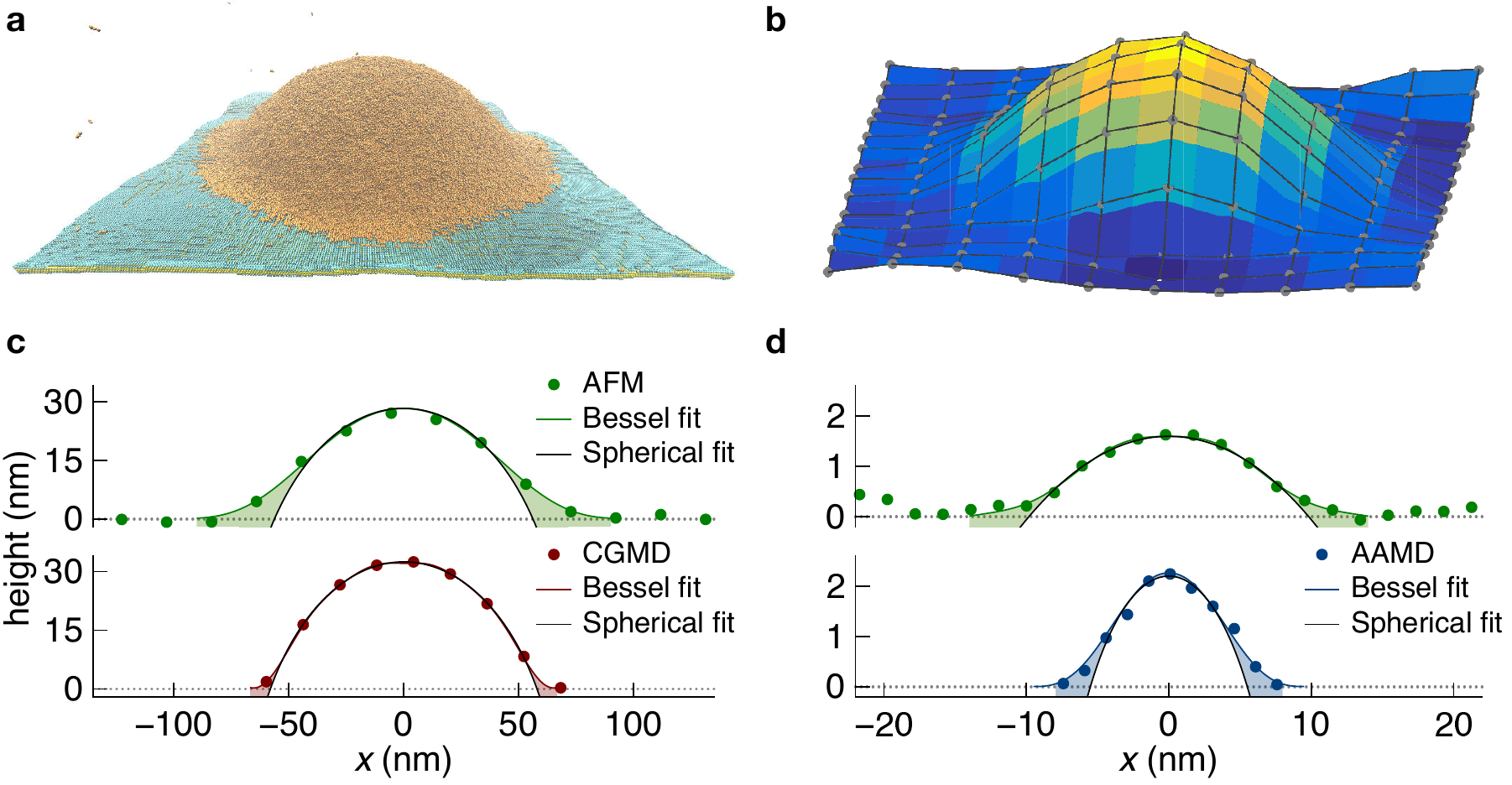}}
\caption{\textbf{Nanoscale droplet topographies: experiment and simulation.} \textbf{a,} 3D representation of a decane droplet obtained from Coarse Grain Molecular Dynamics (CGMD) simulation (height: 30 nm). A decane molecule is represented by 3 orange beads, and light blue beads form the glass surface. For clarity, Nitrogen beads are not shown. \textbf{b,} Measured Atomic Force Microscope (AFM) topography of a decane droplet (height: 30 nm). \textbf{c,} Cross sectional decane droplet profiles obtained from AFM measurements (green dots) and CGMD simulations (red dots), for droplets having heights of 30 nm. \textbf{d,} Decane droplet profiles obtained from AFM measurements (green dots) and All-Atom Molecular Dynamics (AAMD) simulations (blue dots), for droplets having heights of 2 nm.  In \textbf{c} and \textbf{d} the colored lines are cuts through the 3D surface fits. The black lines represent spherical cap fits. The shaded areas highlight the deviation of the droplets' topographies from idealized spherical caps.} 
\label{fig1}
\end{figure*}

To investigate systematically the droplet shape deviations due to molecular scale liquid-solid interactions, we have both measured and simulated decane droplets on glass having volumes ranging from about $10{^2}\,\textnormal{nm}^3$ to $10{^7}\,\textnormal{nm}^3$, spanning five orders of magnitude. By direct comparison with All-Atom Molecular Dynamics (AAMD) and Coarse-Grained Molecular Dynamics (CGMD) simulations, we confirm that we have measured the shapes of droplets of, at the smallest limit, approximately 1000 decane molecules only. In Fig.~\ref{fig1} we plot representative droplet topographies and cross sections for comparison. In Fig.~\ref{fig1}(a) and (b) are a simulated droplet and a measured surface, respectively, for decane droplets with heights of approximately $30\,\textnormal{nm}$. Fig.~\ref{fig1}(c) shows cross sections of the surfaces through the centers of the measured and simulated droplets from (a) and (b). Fig.~\ref{fig1}(d) shows cross sections through measured and simulated droplets with heights of about $2\,\textnormal{nm}$. To extract geometrical information, we fit the three-dimensional measured and simulated droplet surfaces to a sum of Bessel functions of the first kind and obtain satisfactory fits by including four Bessel terms.
From the surface fits, we determine geometrical information, such as height, volume, and contact area for further analysis.

Larger drops of decane on glass, surrounded by air, are indeed well approximated by a spherical shape, and the macroscopic contact angle for any such droplet is measured to be approximately $\theta=4^{\circ}$~\cite{Horng2010}. To compare our measurements at small scales directly with the macroscopic system, we fit a spherical cap to each individual droplet.  From these fitted caps we can extract for each droplet an effective contact angle. In Fig.~\ref{fig1}(c) and (d), cuts through the surface fits are shown as the colored lines that pass through measured and simulated data, and cuts through the fitted spherical caps are shown with black lines. While the Bessel fits capture the droplets' surface features, the spherical caps generally underestimate the non-spherical contributions occurring predominantly in the vicinity of the liquid-solid interface and, for smaller droplets, even at the droplet peaks. The shaded areas highlight the deviations of the actual droplet topographies from spherical caps.

For the larger droplets, pictured in Fig~\ref{fig1}(c), AFM and CGMD reveal similar profiles, with the only significant difference appearing near the glass surface. The smaller droplets in Fig.~\ref{fig1}(d) show larger overall variations, which is expected at this scale, where surface imperfections more strongly affect droplet shapes. Even with the largely different time scales involved for achieving equilibrium conditions in molecular dynamics simulation and AFM measurements, we obtain consistent experimental and simulated droplet shapes and are able to identify the key topographical features in both data sets. Strikingly, the systematic underestimation of droplet contact area and volume by the spherical cap approximation is evident for both experimental and simulated droplets. We note that droplets smaller than 2~nm in height tend to resemble a packaged cluster of molecules over precursor film, or, in other words, a monolayer of decane molecules, see Fig.~\ref{fig4}(c), an effect that we will discuss in more detail below . 

\begin{figure}[htbp]
\centerline{\includegraphics[width=8.0 cm]{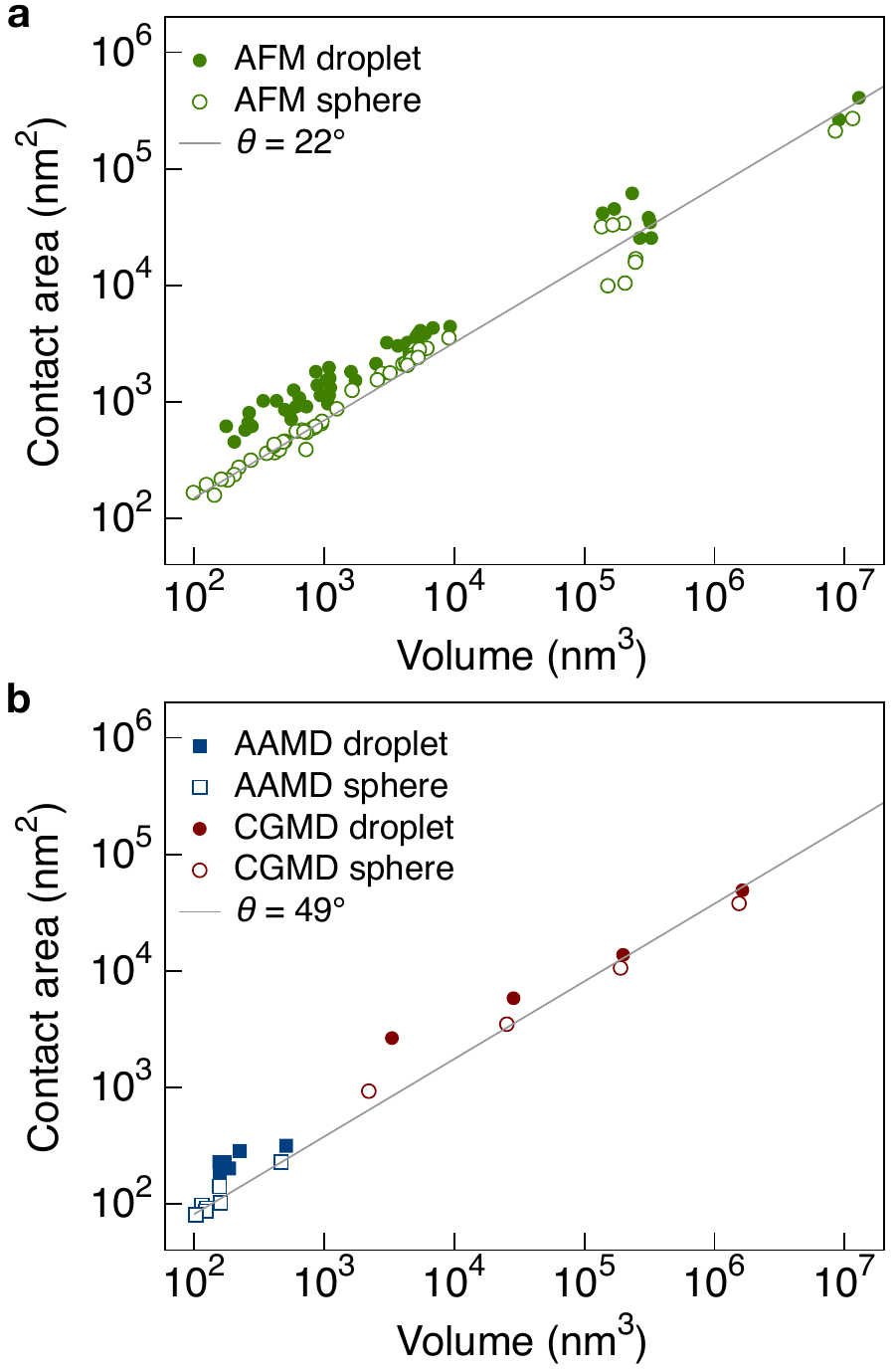}}
\caption{\textbf{Geometric scaling of droplets: systematic deviation from spherical shapes.} \textbf{a,} Droplets measured by AFM. \textbf{b,} Droplets simulated by AAMD and CGMD. Both plots exhibit an offset between surface contact areas of actual droplets and spherical cap approximations of the same droplets, demonstrating a systematic underestimation of droplet contact area in the spherical shape approximation. The offset increases with decreasing droplet volume. Straight lines represent calculated contact areas as a function of volume for spherical caps based on the average contact angles extracted from the spherical cap fits; \textbf{a} $\theta=22^{\circ}$ and \textbf{b}  $\theta=49^{\circ}$.} 
\label{fig2}
\end{figure}

Having both an actual surface fit and a spherical cap approximation for each droplet, we can quantify the degree of deviation between the two representations for various geometrical characteristics. In Fig.~\ref{fig2} we show the contact area of (a) experimental and (b) simulated droplets across five orders of magnitude of volume.  For the filled symbols, the contact area was extracted from the fitted surfaces representing the actual droplet shapes.  The open symbols show the contact area for the spherical cap approximation that was fit individually to each droplet. As a key result, spherical caps systematically  underestimate the contact area. On average, we find that the contact area of a fitted spherical cap is only $53\%$ of the actual contact area for each of the measured and simulated cases. Towards larger droplet volumes, actual shapes and spherical approximations tend to converge, which agrees with our expectations.

Because of the small scales involved, where molecular and surface phenomena are significant, variation in the droplets' shapes is expected\cite{giacomello2016wetting}. This variation is evidenced by the scatter in Fig.~\ref{fig2} and is due to residual chemical and topographical heterogeneity in the local environment of the droplets. Nevertheless, the average values are significantly different from the contact angle for macroscopic systems of $\theta=4^{\circ}$.  For comparison, we plot as solid lines the calculated contact area as a function of volume for spherical caps of constant, averaged contact angles of $\theta=22^{\circ}$ extracted from the measured droplets and of $\theta=49^{\circ}$ as obtained from the simulated droplet data. These deviations from the macroscopic contact angle are caused by line tension effects that increasingly affect droplet shapes as droplet volumes decrease. We will discuss this effect in more detail below and in Fig~\ref{fig4}.

Based on these differences in the measured contact angles across scales, it is clear that a new metric for wettability is needed in order to quantify molecular scale wetting phenomena and to be applicable from the nanoscale to the macroscale. In the following, we will use for this purpose the system's adsorption energy density, $\alpha$, which we define as the difference in energy density between the actual droplet in contact with the surface and a droplet with an equivalent volume or a nearly equivalent number of molecules, but not in contact with the surface. This metric has the benefit that it is experimentally accessible in that one can assume a simple energetic model and calculate $\alpha$ directly from a droplet's shape. Here we include internal bulk energies for solid, liquid, and vapor constituents. We also include surface tension for the liquid-vapor interface, and we assume that the liquid-solid interactions are represented by an effective short-range potential with a sharp cutoff at a constant height above the surface. We calibrate the values of internal and potential energies with $\alpha$ calculated from molecular dynamics simulations, where system energy is directly accessible.

\begin{figure}[htbp]
\centerline{\includegraphics[width=8.5 cm]{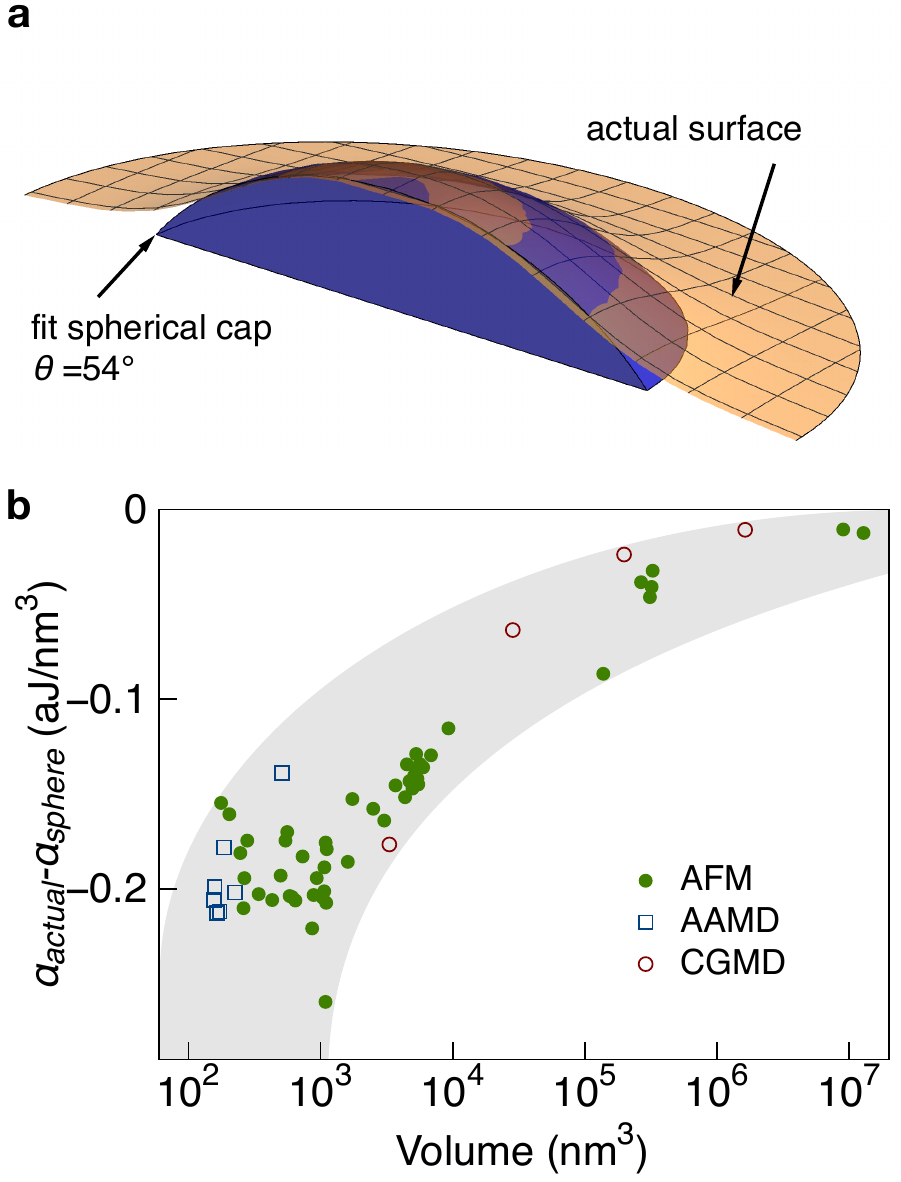}}
\caption{\textbf{Droplet adsorption energy: underestimated at the nanoscale.} \textbf{a,} Comparison of the actual surface with an idealized spherical cap fit to the same data. \textbf{b,} Difference between adsorption energy for the actual surface and that of the spherical cap approximation. The negative differences indicate that the fitted spherical cap underestimates adsorption energy, just as it underestimates contact area (see Fig.~\ref{fig2}).  For volumes larger than $10^6\,\textnormal{nm}^3$ a spherical cap fit provides a robust estimate of the adsorption energy $\alpha$.} 
\label{fig3}
\end{figure}

In Fig.~\ref{fig3} we compare the adsorption energies of the actual droplet shapes to those of the corresponding spherical cap approximations.  For reference, Fig.~\ref{fig3}(a) shows a cut through an actual surface of a simulated droplet and its fitted spherical cap.  Figure~\ref{fig3}(b) shows the difference between $\alpha_{\textnormal{actual}}$ for the actual surface and $\alpha_{\textnormal{sphere}}$ for the idealized spherical cap, for all droplets.  The negative differences in Fig.~\ref{fig3}(b) indicate that the fitted spherical cap underestimates a droplet's adsorption energy.  We note that the degree of underestimation decreases with increasing scale. Ultimately, for volumes above $10^6\,\textnormal{nm}^3$, differences in adsorption energy density vanish, and a spherical cap fit provides a fair approximation of a droplet's actual shape.

\begin{figure}[htbp]
\centerline{\includegraphics[width=15.0 cm]{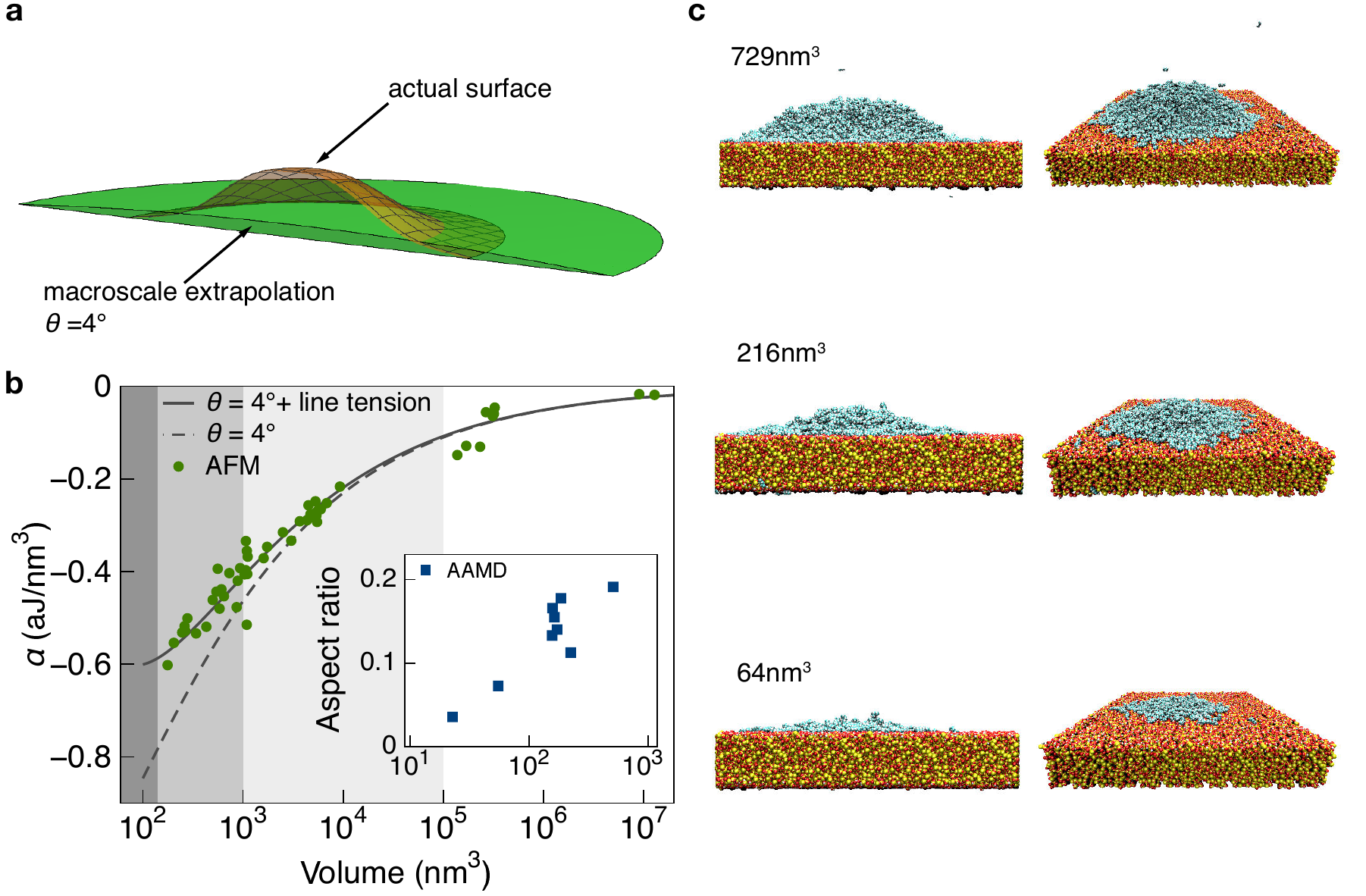}}
\caption{\textbf{Breakdown of spherical cap approximation at the nanoscale.} \textbf{a,} Visualization of an actual nanoscale droplet surface for comparison with a (non-realistic) extrapolation of the macroscopic droplet shape of equivalent volume. \textbf{b,} Adsorption energy of measured droplets (circles) and adsorption energy calculated with the macroscopic model assuming a macroscopic contact angle of $4^{\circ}$ (dashed line) as function of droplet volume. By including a line tension of $4.1\times 10^{-10}\,\textnormal{N}$, the adsorption energy of the macroscopic model is corrected to that of the nanoscale droplets (solid line). In the inset, squares mark the aspect ratio (droplet height/contact diameter) of droplets simulated by All-Atom Molecular Dynamics (AAMD).  The decrease in aspect ratio with decreasing volume reflects the tendency of very small quantities of molecules to form precursor films rather than well-defined droplets.  \textbf{c,} Representative visualizations of molecular arrangements, demonstrating the breakdown of droplet formation in favor of energetically favourable precursor films.}
\label{fig4}
\end{figure}

To demonstrate the compatibility of $\alpha$ with the macroscopic concept of a drop as a spherical cap having a contact angle $\theta=4^{\circ}$, we directly extrapolate such a shape to the nanoscale and calculate its adsorption energy.  For reference, in Fig.~\ref{fig4}(a) we show the actual surface of a nanoscale droplet overlaid with the fictitious, extrapolated drop of the same volume, that has $\theta=4^{\circ}$. The two shapes are quite different, and for our system the extrapolated drop overestimates the contact area and magnitude of $\alpha$, as indicated in Fig.~\ref{fig4}(b) by the dashed line.

Though the macroscopic idealization does not reproduce the actual shape of nanoscale droplets, it is possible to recover the energetic balance of the system by including a line tension term\cite{Amirfazli2004} proportional to the length of the three-phase contact line. As indicated by the solid line in Fig.~\ref{fig4}(b), by including a line tension of $4.1\times10^{-10}\,\textnormal{N}$ in the adsorption energy of the fictitious, macroscopic drops of Fig.~\ref{fig4}(a), $\alpha$ can be calibrated to match the adsorption energy derived from the actual shapes of measured droplets. Note that the line tension value used here agrees with theoretical expectations\cite{liu2013accurate,Checco2003,ward2008effect,Berg2010}.

In the inset of Fig.~\ref{fig4}(b) we show the aspect ratio, i.e. droplet height divided by contact diameter, for the smallest simulated droplets.  
The points reveal how aspect ratio decreases with decreasing volume. On a linear axis, this decrease would appear much more abrupt, and it reflects a tendency of small quantities of molecules to spread into precursor films rather than to form droplets. This departure from a droplet can be seen explicitly in Fig.~\ref{fig4}(c), which shows the molecular arrangements for three droplets with volumes in this region. It is therefore not likely that we can observe, either experimentally or computationally, droplets smaller than those already included in the calculation of $\alpha$. More broadly, we believe that the droplet formation and scaling characteristics reported here should occur in similar systems, but with quantitative differences that are determined by the energetic balance of the system's constituents.

In summary, by rigorously comparing experiment and simulation, we have investigated the affinity of a liquid for a solid surface at the nanoscale. We have demonstrated the breakdown of the spherical cap approximation of a droplet's geometry, calling for the introduction of a metric to quantify wettability, applicable across scales. To that end, we have introduced as a suitable metric the adsorption energy density, which enables proper quantification of a liquid's affinity for a solid surface, even at the nanoscale. We have shown that the macroscopic concept of a drop as a spherical cap can be calibrated to have the correct nanoscale adsorption energy if line tension is included. Future research is aimed at extending this approach to other systems in order to establish it as a consistent metric, benefiting future technological applications of liquid-solid interactions at the nanoscale.

\begin{methods}
\subsection{Materials and AFM measurements}
We prepare decane droplets on glass surfaces by starting with commercially available glass cover slips (Paul Marienfeld GmbH \& Co. KG). We clean the glass surfaces in an ultrasonic bath with acetone followed by isopropanol (Sigma-Aldrich Co. LLC.). Samples are blown dry with clean nitrogen. Then decane (anhydrous, $\geq$99\%, Sigma-Aldrich Co. LLC.) is introduced in an ultrasonic nozzle (Atomizer, Sonics \& Materials, Inc.). Upon vaporization a fine mist of decane is applied to the glass surface. All AFM measurements have been performed with a Dimension 3100 (Veeco) operated in tapping mode with silicon non-contact tips (TESP-V2, Bruker Nano Inc.).

\subsection{All-Atom Molecular Dynamics - AAMD}
The system studied consists of decane droplets on a glass surface, surrounded by air. Because air is 78\% Nitrogen, we approximate it with Nitrogen gas in all numerical simulations. The glass surface was obtained by cutting a slab with thickness of approximately 4.2~nm from the bulk. The glass bulk was created using the Amorphous Silicon Dioxide builder from the VMD package\cite{Hump96}. The dangling bonds from Silicon and Oxygen atoms were passivated with hydroxyl (OH) and Hydrogen, respectively, resulting in a hydroxylated glass surface. The concentration of hydroxyl was 5.0/nm$^2$, close to the experimental value of 4.9/nm$^2$~\cite{Zhuravlev2000}. In the next step a cubic box containing decane molecules surrounded by Nitrogen molecules was placed over the passivated glass slab. The number of decane and Nitrogen molecules and the volume of the simulation box were chosen to reproduce the equilibrated density of pure Nitrogen and pure decane, obtained through a set of simulations starting from the NVE ensemble, then the NVT, and finally the NPT ensemble (for more details see supplementary information). For the glass/decane/Nitrogen system we performed AAMD simulations with the NVT ensemble only. In all simulations, periodic boundary conditions were used in all directions. In the AAMD simulations the interactions between atoms were modeled using a classical force field. For decane, CHARMM-based force field\cite{Brooks2009} parameters were used. For Nitrogen the Lennard-Jones potential plus point-charge models\cite{Potoff2001} were used. For the simulation of the hydroxylated glass slab we used the CHARMM Water Contact Angle (CWCA) force field\cite{CruzChu2006}. To account for long-range electrostatic interactions, the reciprocal space Particle-particle Particle-mesh (PPPM) method\cite{hockney1988computer} was adopted. For all calculations we used a time step of 0.5~fs, a cutoff radius of 1~nm for van der Waals and Coulomb interactions, a temperature of 300~K, and a pressure of 1~atm. To control temperature and pressure, Nose-Hoover thermostats and barostats were used with a relaxation time of 0.1~ps and 1~ps, respectively. All MD simulations were performed with the LAMMPS package\cite{Plimpton1995}.

\subsection{Coarse-Grained Molecular Dynamics - CGMD}
CGMD is based on an effective description of the system with reduced degrees of freedom. The typical approach is to represent a set of atoms with a single ``super-atom,'' which we refer to in the following as a CG bead. The most intuitive way is to represent CG beads as the center of mass of a set of atoms. For glass we define two types of beads: surface beads and bulk beads. A bulk bead represents three Silicon and six Oxygen atoms, while a surface bead represents three Silicon, six Oxygen, and two Hydrogen atoms. Decane molecules are represented as three beads and Nitrogen molecules as one single bead. The interaction potentials of CG beads are a priori unknown and must be determined by using a numerical routine based on AAMD simulations. In the present case, the systematic, iterative Boltzmann inversion (IBI) numerical scheme from Fukuda \textit{et al.}\cite{Fukuda2013} was adopted. In this method, effective pair potentials between CG beads are determined from AAMD with an iterative refinement procedure. In this way, the CG bond length and bond angle potentials for glass beads, and the CG nonbonded potentials for Nitrogen beads, Nitrogen/glass surface beads and decane/glass surface beads, were obtained. For the interaction between decane beads, the MARTINI coarse-grained force field was used\cite{Marrink2007}. The coarse-grained glass surface was built with three layers of bulk glass beads, and two layers, on top and at the bottom, of glass surface beads. The same procedure as in AAMD was followed to build and simulate the glass/decane/Nitrogen system. The only difference was in the parameters: time step of 5 fs and relaxation time for Nose-Hoover thermostats and barostats of 0.25 and 2.5 ps, respectively.

\subsection{Surface Fits}
The shapes of the actual surfaces of simulated and measured droplets were extracted by fitting droplet surface data to a partial expansion in Bessel functions of the first kind, $J_i$:
\begin{equation}
C + \sum_{i=0}^4 c_i J_i \big(k(\theta)r\big),
\end{equation}
with $c_1=0$. To treat the break in cylindrical symmetry due to the AFM scan axis, we scale the radius by 
\begin{equation}
k(\theta) = \frac{\sqrt{(a \sin\theta)^2+(b\cos\theta)^2}}
{a b},
\end{equation}
where $\theta=0$ aligns with the scan axis. Fit parameters are the constant offset, $C$, the expansion coefficients $c_i$, and the radial scaling factors $a$ and $b$. With four non-zero Bessel terms we find good fits to all droplets.

\subsection{Effective potential}
The adsorption energy density, $\alpha$, is defined as the difference in energy density between the droplet on the surface and the droplet away from the surface. The first term of the difference was obtained from MD by collecting the total internal energy of the system and then dividing by the droplet volume. To keep the decane droplet from interacting with or adsorbing on the surface, the second term of the difference was obtained in a different way. A spherical decane droplet was put inside a cubic box and surrounded by Nitrogen molecules, and the same procedure described above was followed. In both configurations, the droplet on the surface and the isolated droplet, the same number of decane molecules was used. Because the two configurations are not identical, we evaluate the difference in energy density by using the superposition principle:
\begin{equation}\label{eq:mdalpha}
\alpha = \frac{(E_\textnormal{DNG}-E_\textnormal{G}-E_\textnormal{N})^\textnormal{on \ surface} - (E_\textnormal{DN}-E_\textnormal{N})^\textnormal{off \ surface}}{V},
\end{equation}
where $V$ is the droplet volume, $E_\textnormal{DNG}$, $E_\textnormal{DN}$, $E_\textnormal{G}$ and $E_\textnormal{N}$ are the total internal energy of the decane/Nitrogen/glass, decane/Nitrogen, and isolated glass and Nitrogen subsystems, respectively. The superscripts ``on surface'' and ``off surface'' refer to the decane/Nitrogen/glass system with a droplet on the glass surface and to the decane/Nitrogen system with an isolated droplet, respectively.

To calculate $\alpha$ for measured droplets, where system energy must be inferred from droplet geometry, we assume thermodynamical equilibrium and use a coarse but simple model where, after cancelling terms,
\begin{equation}
\alpha = G_{\textnormal{DG}}\frac{V_p^{\textnormal{on surface}}}{V}
+\lambda_{\textnormal{DNG}}\frac{P\,^{\textnormal{on surface}}}{V}
+U_{\textnormal{DN}}\frac{SA^{\textnormal{on surface}}-SA^{\textnormal{off surface}}}{V}.
\end{equation}
The first term on the right hand side includes the contribution from an effective potential energy of strength $G_{\textnormal{DG}}$, which we use to represent the interaction between decane molecules and the glass surface.  For simplicity we assume it to have a constant value extending from the surface to a cutoff height, $z_{\textnormal{cutoff}}$.  The portion of a droplet's volume that is within this region is denoted $V_p$ and is extracted from surface fits or spherical cap fits, as appropriate.  The second term contains the contribution from line tension and is only applied to the extrapolated macroscale drop with contact angle $4^\circ$ (see Fig.~\ref{fig4}). The three-phase line tension is $\lambda_{\textnormal{DNG}}$, and $P\,^{\textnormal{on surface}}$ is the perimeter of the contact area. In the third term, $U_{\textnormal{DN}}$ is the surface tension of the decane/Nitrogen subsystem.  Surface areas are denoted $SA$, and we extract $SA^{\textnormal{on surface}}$ from the surface fits or spherical cap fits, while $SA^{\textnormal{off surface}}$ is for a sphere of the same volume.

We use the tabulated value\cite{jasper1955isobaric} of $U_{\textnormal{DN}}=0.0238\,\textnormal{aJ}/\textnormal{nm}^2$.
By choosing $G_{\textnormal{DG}}$ and $z_{\textnormal{cutoff}}$ to be $-1.3\,\textnormal{aJ}/\textnormal{nm}^3$ and $0.2\,\textnormal{nm}$, respectively, we reproduce well for MD droplets the adsorption energy density calculated in eq.~(\ref{eq:mdalpha}), with values taken directly from MD simulation.  Line tension is only applied to the extrapolated macroscale drops (see Fig.~\ref{fig4}), where to match values calculated from actual droplet shapes, we need $\lambda_{\textnormal{DNG}}=4.1\times 10^{-10}\,\textnormal{N}$.
This value of line tension is within the range of magnitudes predicted theoretically\cite{liu2013accurate,Checco2003,ward2008effect,Berg2010}.  The numerical value depends on the details of our model, which we created specifically to be calibrated and then to depend only on droplet topography.  A more precise value may be obtained using a more detailed thermodynamical model.

\end{methods}

\begin{addendum}
 \item[Acknowledgements] We acknowledge discussion with and support by Ulisses Mello, Claudius Feger, Phaedon Avouris (IBM Research), Rafael R. Del Grande (UFRJ-Rio de Janeiro), Alexandre A. L. Cunha, William F. L. Candela, José A. F. Gutiérrez, Marcio S. Carvalho (PUC-Rio de Janeiro), Laura P. C. Amorim, Ado Jorio, Glaura Goulart Silva (UFMG-Belo Horizonte), Carlos Speglich, Lua Selene S. Almeida (CENPES-Rio de Janeiro), and Tito Bonagamba (USP-São Carlos).
 \item[Correspondence] Correspondence and requests for materials
should be addressed to Ronaldo Giro~(email: rgiro@br.ibm.com) and Mathias Steiner~(email: msteine@us.ibm.com) .
 
\end{addendum}

\clearpage
\section{Supplementary Methodology}
\subsection{All-Atom Molecular Dynamics (AAMD)}

We simulated in three dimensions the behavior of a decane droplet surrounded by nitrogen gas over a glass surface at the molecular scale by using AAMD. 

\begin{figure}[htbp]
\centerline{\includegraphics[width=10.0 cm]{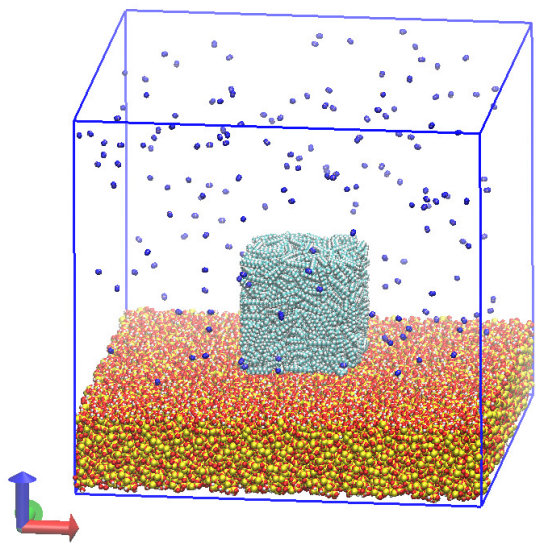}}
\caption{(a) 3D representation of the input configuration of decane molecules (light blue) packed in a cubic arrangement over a glass surface (yellown and red) and surrounded by nitrogen molecules (dark blue).} 
\label{fig1-supp}
\end{figure}

A set of cubic boxes composed of decane molecules and surrounded by nitrogen molecules was placed onto a passivated glass slab (see Table~\ref{tab1} for details). The molecules were distributed randomly in defined regions of the simulation box, as shown in Fig.~\ref{fig1-supp}, and packaged by using the Packmol package\cite{Martinez2009}.

The dimensions of the simulation boxes for the system consisting of glass, decane, and nitrogen, are shown in Table~\ref{tab1}. The dimensions labeled L were chosen in order to isolate the interactions between periodic images. Periodic boundary conditions were used in all directions. The number of decane and nitrogen molecules, as well as the volume of the simulation box, were chosen in order to reproduce the equilibrated density of pure nitrogen and pure decane, obtained by following the protocol for a specific temperature and pressure: a 1 ps run was performed for the NVE ensemble (micro canonical ensemble - constant Number of atoms, Volume, and Energy), followed by 10 ps for the NVT ensemble (canonical ensemble - constant Number of atoms, Volume, and Temperature), and finally 2 ns for the NPT ensemble (isothermal-isobaric or grand canonical ensemble - constant Number of atoms, Pressure, and Temperature). The protocols were performed for a cubic box (10 nm edge) containing 3089 decane molecules, and for a cubic box (20 nm edge) containing 160 nitrogen molecules. In both cases periodic boundary conditions were used in all three dimensions. The preliminary CMD simulations were performed first for decane and the isolated nitrogen systems, because the density of the glass/decane/nitrogen system cannot be equilibrated using the NPT ensemble, as the NPT ensemble does not contain enough nitrogen molecules to transfer energy from collisions (equivalent to pressure) to the glass surface. As a consequence, in the case of the glass/decane/nitrogen system, we performed CMD simulations with the NVT ensemble only.

\begin{table}[htbp]
\centering
\caption{Input box sizes and number of molecules considered in AAMD simulations.}
\label{tab1}
\begin{tabular}{cccccc}
\hline \hline
decane cubic & \# decane & \# nitrogen & \multicolumn{3}{c}{System box size} \\ 
box side         & molecules    & molecules      & L$_x$ (nm)  & L$_y$ (nm)   & L$_z$ (nm)  \\ \hline
6.0              & 670          & 180            & 23         & 23         & 21.7      \\
6.1              & 704          & 180            & 23         & 23         & 21.7      \\
6.2              & 739          & 180            & 23         & 23         & 21.7      \\
6.3              & 776          & 180            & 23         & 23         & 21.7      \\
6.4              & 813          & 180            & 23         & 23         & 21.7      \\
7.0              & 1045         & 370            & 28.5       & 28.5       & 25.6      \\
9.0              & 2257         & 316            & 28.5       & 28.5       & 29.5      \\ 
13.0             & 6815         & 1030           & 40.1       & 40.1    & 34.5      \\ \hline \hline 
\end{tabular}
\end{table}

\subsection{Coarse-Grained Molecular Dynamics (CGMD)}
\begin{figure}[htbp]
\centerline{\includegraphics[width=15.0 cm]{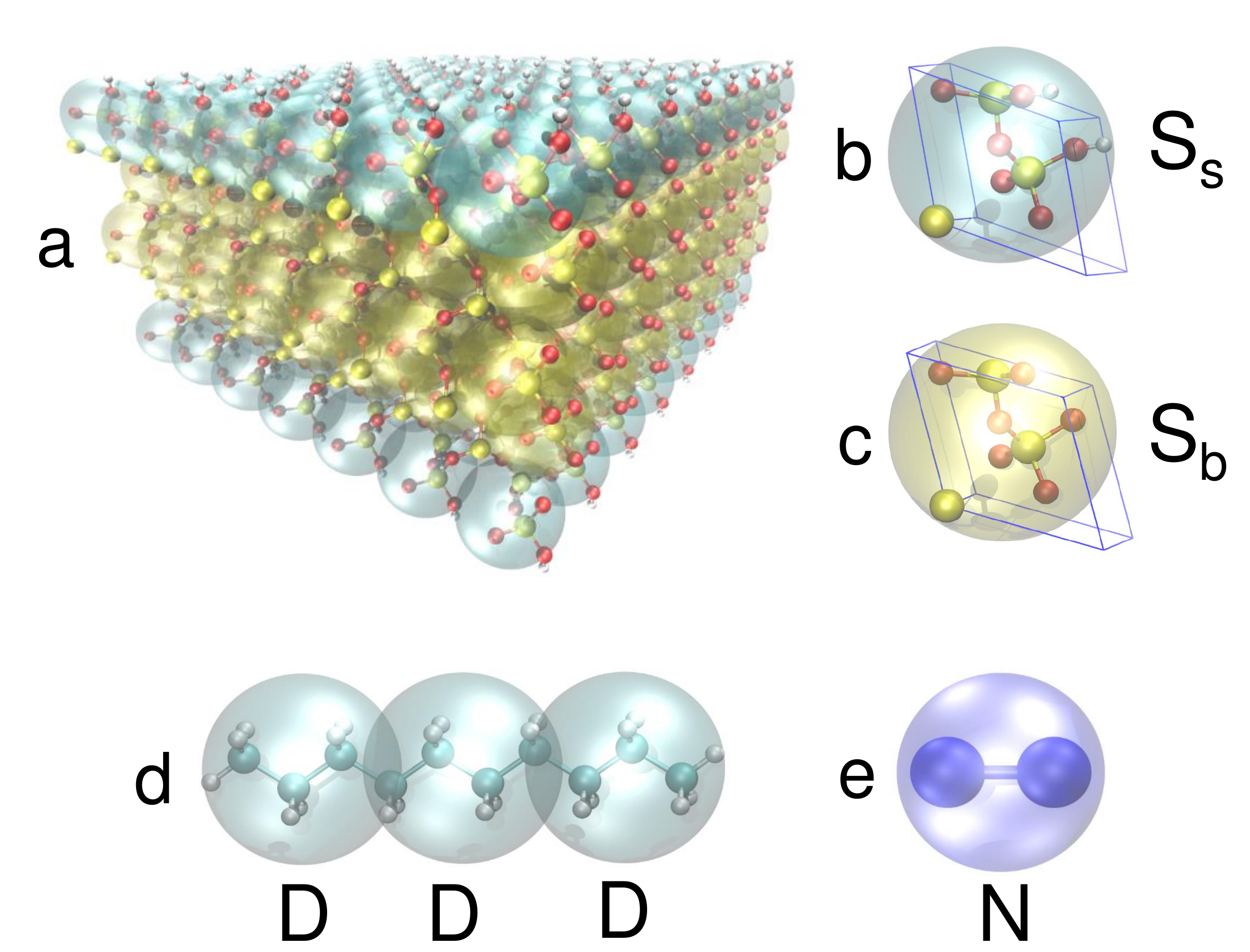}}
\caption{Coarse-grained (transparent) and All-Atom representation of (a) glass surface with three bulk layers (in transparent yellow) and two surface layers (in transparent light blue). (b) and (c) are the unit cells of a surface and a bulk bead. (d) and (e) are the CG representations of decane and nitrogen, respectively. In the All-Atom representation, Silicon atoms are in yellow, Oxygen in red, Hydrogen in white, Carbon in light blue, and Nitrogen in dark blue. The capital letters, S$_\textnormal{b}$, S$_\textnormal{s}$, D and N, label the coarse-grained beads types.}  
\label{fig2-supp}
\end{figure}

In CGMD a set of atoms is represented as a single ``super-atom,'' which we will refer to as CG bead. The most intuitive procedure is to represent CG beads as the center of mass of a set of atoms.  For glass we define two types of beads:  surface beads and bulk beads.  A bulk bead represents three Silicon and six Oxygen atoms, while a surface bead represents three Silicon, six Oxygen, and two Hydrogen atoms. Decane molecules are represented as three beads, and Nitrogen molecules as one single bead.  The coarse-grained glass surface was built with three layers of bulk glass beads, and two layers, one on top and one on the bottom, arranged in a deformed cubic lattice (see Fig.~\ref{fig2-supp} for details). In Table~\ref{tab2} we show the lattice vectors and the All-Atom unit cell coordinates for surface and bulk glass CG beads.

\begin{table}[htbp]
\centering
\caption{Lattice vectors and the All-Atom unit cell coordinates for surface and bulk glass CG beads. Dimensions in Angstroms. }
\label{tab2}
\begin{tabular}{lccc}
\multicolumn{4}{c}{primitive lattice vectors} \\ \hline \hline
               & $\hat{x}$            & $\hat{y}$             &$\hat{z}$            \\ \hline
$\vec{a}$              & 5.400        & 0.000         & 0.000        \\
$\vec{b}$              & 0.000        & 4.910         & 0.000        \\
$\vec{c}$              & 0.000        & -2.455        & 4.252        \\ \hline \hline
               &              &               &              \\
\multicolumn{4}{c}{atomic coordinates - CG unit cell}         \\ \hline \hline
               & x            & y             & z            \\ \hline
               & 2.281        & 1.761         & 1.796        \\ \hline \hline
               &              &               &              \\
\multicolumn{4}{c}{atomic coordinates - AA surface unit cell}    \\ \hline \hline
chemical element        & x            & y             & z            \\ \hline
Si             & 0.000        & 0.000         & 0.000        \\
Si             & 3.601        & 1.086         & 2.208        \\
Si             & 1.800        & 3.541         & 2.045        \\
O              & 1.171        & 4.028         & 0.639        \\
O              & 2.430        & 2.081         & 1.763        \\
O              & 2.972        & -0.374        & 2.489        \\
O              & 4.772        & 0.975         & 1.124        \\
O              & 4.230        & 1.573         & 3.614        \\
O              & 0.629        & 3.430         & 3.128        \\
H              & 4.230        & 1.573         & 4.514        \\
H              & 0.629        & 3.430         & 4.028        \\ \hline \hline
               &              &               &              \\
\multicolumn{4}{l}{atomic coordinates - AA bulk unit cell}    \\ \hline \hline
chemical element        & x            & y             & z            \\ \hline
Si             & 0.000        & 0.000         & 0.000        \\
Si             & 3.601        & 1.086         & 2.208        \\
Si             & 1.800        & 3.541         & 2.045        \\
O              & 1.171        & 4.028         & 0.639        \\
O              & 2.430        & 2.081         & 1.763        \\
O              & 2.972        & -0.374        & 2.489        \\
O              & 4.772        & 0.975         & 1.124        \\
O              & 4.230        & 1.573         & 3.614        \\
O              & 0.629        & 3.430         & 3.128        \\ 
\hline \hline
\end{tabular}
\end{table}

The CG interaction potential was built following a mapping procedure from AAMD. Due to difficulties in mapping amorphous materials such as glass, we considered $\alpha$-quartz in (100) orientation. We expect this simplification to be a good approximation because: (i) The glass surface beads represent an average interaction of three Silicon, six Oxygen, and 2 Hydrogen atoms mapped from $\alpha$-quartz AAMD. This average over a set of atoms results in a certain indistinguishability among the atoms' arrangements (amorphous in glass and crystalline in $\alpha$-quartz). (ii) The glass surface beads mapped from All-Atom $\alpha$-quartz shows the same concentration of hydroxyl (OH) as glass, i.e around 5 hydroxyls per nm$^2$. Thus the same surface chemistry as glass, is characterized. 

By mapping $\alpha$-quartz from All-Atom to the CG representation we extracted the bond length and bond angle potentials for glass, and the CG nonbonded potentials to describe the interactions between Nitrogen and decane with glass surface beads. In the present case, the systematic iterative Boltzmann inversion (IBI) numerical scheme from Fukuda \textit{et al.}\cite{Fukuda2013} was adopted. In this method, effective pair potentials between CG beads are determined from AAMD with an iterative refinement procedure.

The bond length potentials are represented by a harmonic potential:
\begin{equation}
U_\textnormal{bond}(R)=K_\textnormal{bond}(R-R_0)^2
\label{eqbond}
\end{equation}

In Table~\ref{tab3} we show the parameters for silica (glass) and decane. The decane parameters are extracted from the MARTINI Force Field\cite{Marrink2007}.

\begin{table}[htbp]
\centering
\caption{Bond length potential parameters for glass and decane beads.}
\label{tab3}
\begin{tabular}{lll}
\hline \hline
bonds between bead types &$K_\textnormal{bond}$ (Kcal/mol\AA$^2$) & $R_0$ (\AA) \\ \hline
S$_\textnormal{b}$ -- S$_\textnormal{b}$               & 125.0                                       & 5.13       \\
S$_\textnormal{b}$ -- S$_\textnormal{b}$               & 105.0                                       & 5.64       \\
S$_\textnormal{b}$ -- S$_\textnormal{s}$               & 75.0                                        & 5.13       \\
S$_\textnormal{s}$ -- S$_\textnormal{s}$               & 105.0                                       & 5.13       \\
S$_\textnormal{s}$ -- S$_\textnormal{s}$               & 47.0                                          & 5.64       \\
D -- D$^a$                     & 1.4938                                      & 4.7 \\
\hline \hline
%\multicolumn{3}{l}{$^a$ parameters from the MARTINI Force Field\cite{Marrink2007}}
\end{tabular}
\end{table}

For glass beads, the bond angle potential is represented by a harmonic potential:
\begin{equation}
U_\textnormal{angle1}(\theta)=K_\textnormal{angle1}(\theta-\theta_0)^2
\label{eqangle1}
\end{equation}

According to Marrink \textit{et al.}\cite{Marrink2007}, the bond angle potential for decane beads is represented as a cosine-squared potential:
\begin{equation}
U_\textnormal{angle2}(\theta)=K_\textnormal{angle2}[\cos(\theta)-\cos(\theta_0)]^2
\label{eqangle2}
\end{equation}

In Table~\ref{tab4} we show the parameters for bond the angle potential for glass and decane beads.

\begin{table}[htbp]
\centering
\caption{Bond angle potential parameters for glass and decane beads.}
\label{tab4}
\begin{tabular}{lccl}
\hline \hline
angles between  & $K_\textnormal{angle1}$             & $\theta_0$    & Potential             \\
bead types                           & (Kcal/mol radian$^2$) & (degrees) &                       \\ \hline
S$_\textnormal{b}$ -- S$_\textnormal{b}$ -- S$_\textnormal{b}$                 & 30.0                & 180       & harmonic -- eq. \ref{eqangle1}       \\
S$_\textnormal{b}$ -- S$_\textnormal{b}$ -- S$_\textnormal{b}$                 & 19.6                & 60        & harmonic -- eq. \ref{eqangle1}       \\
S$_\textnormal{b}$ -- S$_\textnormal{b}$ -- S$_\textnormal{b}$                 & 47.6                & 90        & harmonic -- eq. \ref{eqangle1}       \\
S$_\textnormal{b}$ -- S$_\textnormal{b}$ -- S$_\textnormal{s}$                 & 15.0                & 60        & harmonic -- eq. \ref{eqangle1}       \\
S$_\textnormal{b}$ -- S$_\textnormal{b}$ -- S$_\textnormal{s}$                 & 50.2                & 90        & harmonic -- eq. \ref{eqangle1}       \\
S$_\textnormal{s}$ -- S$_\textnormal{b}$ -- S$_\textnormal{s}$                 & 14.5                & 60        & harmonic -- eq. \ref{eqangle1}      \\
S$_\textnormal{s}$ -- S$_\textnormal{s}$ -- S$_\textnormal{s}$                 & 36.6                & 90        & harmonic -- eq. \ref{eqangle1}       \\
S$_\textnormal{s}$ -- S$_\textnormal{s}$ -- S$_\textnormal{s}$                 & 16.0                & 180       & harmonic -- eq. \ref{eqangle1}       \\ \hline
angles between  & $K_\textnormal{angle2}$             & $\theta_0$    & Potential             \\
bead types                           & (Kcal/mol)          & (degrees) &                       \\ \hline
D--D--D$^a$                    & 2.9876              & 180       & cosine squared -- eq. \ref{eqangle2} \\
\hline \hline
\multicolumn{4}{l}{$^a$ parameters from the MARTINI Force Field\cite{Marrink2007}}
\end{tabular}
\end{table}

To describe the nonbonded interactions between Nitrogen-Nitrogen, Decane-Surface glass beads, and Nitrogen-Surface glass beads, a shifted Lenard-Jones (LJ) 12-6 potential energy function was employed as an initial guess for the IBI method: 
\begin{equation}
U_\textnormal{LJ}(r)=4\epsilon _{ij}\left[\left ( \frac{\sigma_{ij} }{r} \right )^{12} - \left ( \frac{\sigma_{ij} }{r} \right )^{6} \right]
\label{eqlj}
\end{equation}

We used the same procedure as described in the work of Fukuda \textit{et al.}\cite{Fukuda2013}, to obtain the parameters $\sigma_{ij}$ and $\epsilon_{ij}$ for the initial guess LJ 12-6 potential. We observed a good agreement between the AAMD and CGMD radial distribution functions and the numerical density profiles for Nitrogen-Nitrogen and Nitrogen-Surface glass beads nonbonded interactions, respectively. Given the good agreement there was no need to proceed with IBI refinement. Thus the nonbonded interactions between these beads are described simply by the LJ 12-6 potential from eq.~\ref{eqlj}.

For Decane-Decane beads, the nonbonded interactions are described by the LJ 12-6 potential, as in eq.~\ref{eqlj}, and the paramenters were obtained from the MARTINI Force Field\cite{Marrink2007}.

In Table~\ref{tab5} we show the $\sigma_{ij}$ and $\epsilon_{ij}$ parameters for nonbonded interactions between Nitrogen-Nitrogen, Nitrogen-Surface glass beads, and Decane-Decane beads.

\begin{table}[htbp]
\centering
\caption{LJ 12-6 potential parameters to describe the unbonded interactions between Nitrogen-Nitrogen, Nitrogen-Surface glass beads, and Decane-Decane beads.}
\label{tab5}
\begin{tabular}{llll}
\hline \hline
unbonded interactions & $\epsilon_{ij}$ (Kcal/mol) & $\sigma_{ij}$ (\AA) & cutoff (\AA) \\
between bead types   &                     &              &              \\ \hline
S$_\textnormal{s}$ -- N                  & 0.57880             & 4.987        & 12.0         \\
D -- D$^a$                   & 0.83596             & 4.700        & 12.0         \\
D -- N                   & 0.41083             & 4.200        & 12.0         \\
N -- N                   & 0.20190             & 3.700        & 12.0  \\ \hline \hline 
\multicolumn{4}{l}{$^a$ parameters from the MARTINI Force Field\cite{Marrink2007}}
\end{tabular}
\end{table}

\begin{figure}[htbp]
\centerline{\includegraphics[width=15.0 cm]{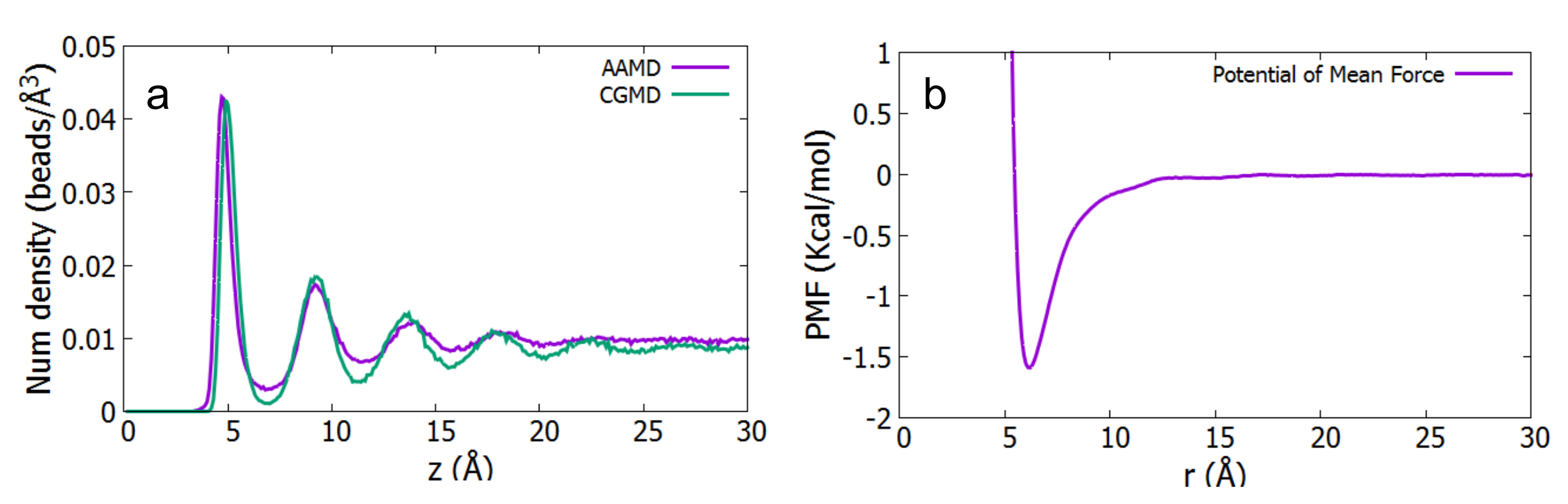}}
\caption{(a) Numerical density profile for decane beads of the glass surface. The numerical density profiles are calculated across the interface between glass and decane fluid. (b) Potential of Mean Force obtained after the IBI refinement procedure.}  
\label{fig3-supp}
\end{figure}

On the other hand, for nonbonded interactions between Decane-Surface glass beads, a good agreement between the numerical density profile from AAMD and CGMD was not observed. Thus the IBI refinement procedure, as described in Fukuda \textit{et al.}\cite{Fukuda2013}, was performed. After the IBI refinement, a good agreement between AAMD and CGMD numerical was obtained (see Fig.~\ref{fig3-supp}(a)).

In Fig.~\ref{fig3-supp}(b) we show the Potential of Mean Force obtained from IBI refinement. We observe a small deviation from the initial guess LJ 12-6 potential in the region from 10 to 15 \AA.

To build the Coarse-Grained system composed of decane over a glass surface and surrounded by nitrogen, the same procedure as described in AAMD section was followed. The dimensions of the simulation boxes are shown in Table~\ref{tab6}.

\begin{table}[htbp]
\centering
\caption{Input box sizes and number of molecules considered in CGMD simulations.}
\label{tab6}
\begin{tabular}{cccccc}
\hline \hline
decane cubic & \# of decane & \# of nitrogen & \multicolumn{3}{c}{System box size} \\ 
box side         & molecules    & molecules      & L$_x$ (nm)  & L$_y$ (nm)   & L$_z$ (nm)  \\ \hline
15.0              & 10450          & 54800            & 156.8         & 157.0         & 106.0      \\
30.0              & 83400          & 54250            & 156.8         & 157.0         & 106.0      \\
57.0              & 591657          & 50365            & 156.8         & 157.0         & 106.0      \\
113.5              & 4519850          & 404600            & 312.9         & 313.4         & 201.3      \\
\hline \hline 
\end{tabular}
\end{table}

\end{document}